\documentclass[12pt,letter,doublespace]{article}
\usepackage[brazil]{babel}
\usepackage{amssymb}
\usepackage{amsmath} 
\usepackage{graphics}
\usepackage{epsfig}
\usepackage{graphicx}
\usepackage{cite}
\usepackage{amsthm}
\usepackage{color}
\usepackage{booktabs}
\usepackage{bbm}
\usepackage{caption}

\usepackage{harvard}

\usepackage{floatrow}
\newfloatcommand{capbtabbox}{table}[][\FBwidth]
\usepackage{blindtext}

\newcommand{\dkonv}{\stackrel{d}{\rightarrow}}

\newcommand{\bqa}{\begin{eqnarray*}}
\newcommand{\eqa}{\end{eqnarray*}}
\newcommand{\bqan}{\begin{eqnarray}}
\newcommand{\eqan}{\end{eqnarray}}
\newcommand{\bit}{\begin{itemize}}
\newcommand{\eit}{\end{itemize}}
\newcommand{\ben}{\begin{enumerate}}
\newcommand{\een}{\end{enumerate}}
\newcommand{\beq}{\begin{equation}}
\newcommand{\eeq}{\end{equation}}
\newcommand{\bdes}{\begin{description}}
\newcommand{\edes}{\end{description}}

\begin{document}
\bibliographystyle{dcu} 
\citationstyle{dcu}

\title{A Review of Kernel Density Estimation with Applications to Econometrics}
\author{\sc Adriano Z. Zambom and Ronaldo Dias\\
Universidade Estadual de Campinas}
\date{December, 2012}
\maketitle{}
\section*{Abstract} Nonparametric density estimation is of great importance when econometricians want to model the probabilistic or stochastic structure of a data set. This comprehensive review summarizes the most important theoretical aspects of kernel density estimation and provides an extensive description of classical and modern data analytic methods to compute the smoothing parameter. Throughout the text, several references can be found to the most up-to-date and cut point research approaches in this area, while econometric data sets are analyzed as examples. Lastly, we present SIZer, a new approach introduced by \citeasnoun{ChaudhuriMarron2000}, whose objective is to analyze the visible features representing important underlying structures for different bandwidths.

\textbf{Keywords:} nonparametric density estimation; SiZer; plug-in bandwidth selectors; cross-validation; smoothing parameter.


\pagestyle{plain}
\setcounter{page}{1}
\setlength{\textheight}{9.0in}
\setlength{\topmargin}{-0.5in}

\section{Introduction}  \label{Sec.Intro}

\indent \indent The field of econometrics focuses on methods that address the probabilistic or stochastic phenomena involving economic data. Modeling the underlying probabilistic structure of the data, i.e., the uncertainty of the process, is a crucial task, for it can be used to describe the mechanism from which the data was generated. Thus, econometricians have widely explored density estimation, both the parametric and nonparametric approaches, to identify these structures and then make inferences about the unknown "true models". A parametric model assumes that the density is known up to a finite number of parameters, while a nonparametric model allows great flexibility in the possible form, usually assuming that it belongs to some infinite collection of curves (differentiable with square integrable second derivatives for example). The most used approach is kernel smoothing, which dates back to \citeasnoun{Rosenblatt1956} and \citeasnoun{Parzen1962}. The aim of this paper is to review the most import aspects of kernel density estimation, both traditional approaches and modern ideas.

\indent A large extent of econometric research concerning estimation of densities has shown that a well estimated density can be extremely useful for applied purposes. An interesting comprehensive review of kernel smoothing and its applications can be found in \citeasnoun{Bierens1987}.
 \citeasnoun{Silverman1986} and \citeasnoun{Scott1992} discuss kernel density estimation thoroughly, giving details about assumptions on the kernel weight, properties of the estimator such as bias and variance, and discusses how to choose the smoothness of the estimate. The choice of the smoothing parameter is a crucial issue in nonparametric estimation, and will be discussed in detail in Section \ref{sec.bandwidth}.

\indent The remainder of this paper is as follows. In Section \ref{sec.histogram} we describe the most basic and intuitive method of density estimation: the histogram. Then, in Section \ref{sec.kernel} we introduce kernel density estimation and the properties of estimators of this type, followed by an overview of old and new bandwidth selection approaches in Section \ref{sec.bandwidth}. Finally, SiZer, a modern idea for accessing features that represent important underlying structures through different levels of smoothing, is introduced in Section \ref{sec.sizer}.

\section{The Histogram} \label{sec.histogram}
\indent \indent The grouping of data in the form of a frequency histogram is a classical methodology that is intrinsic to the foundations of a variety of estimation procedures. Providing useful visual information, it has served as a data presentation device, however, as a density estimation method, it has played a fundamental role in nonparametric statistics.

\begin{figure}[!htbp] 
	\centering
	\includegraphics[width=\textwidth]{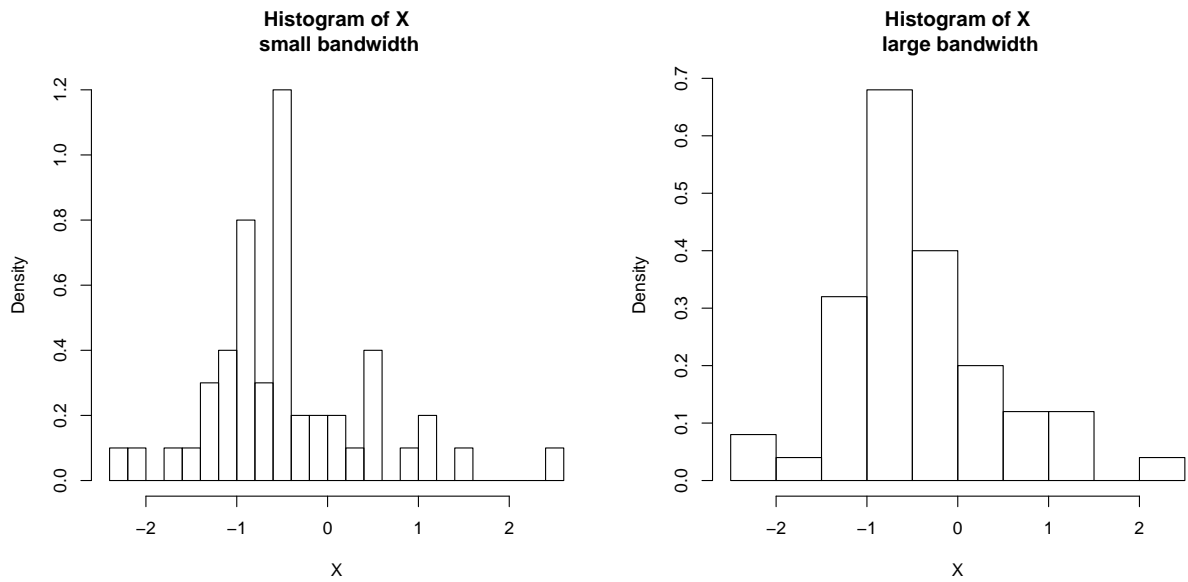}
	\caption{Histogram estimate with small bandwidth (left) and large bandwidth (right)} \label{Fig.hist}
\end{figure}

Basically, the histogram is a step function defined by bin heights, which equal the proportion of observations contained in each bin divided by the bin width. The construction of the histogram is very intuitive, and to formally describe this construction, we will now introduce some notation. Suppose we observe random variables $X_1, \ldots, X_n$ i.i.d. from the distribution function $F_X$, and that $F_X$ is absolutely continuous with respect to a Lesbegue measure on $\mathbb{R}$. Assume that $x_1, \ldots, x_n$ are the data points observed from a realization of the random variables $X_1, \ldots, X_n$. Define the bins as $I_j = [x_0 + jh, x_0 + (j+1)h), j = 1,\ldots,k$, for a starting point $x_0$. Note that
\bqan \label{eq.f}
P(X \in I_j) = \int_{I_j}f(x)dx = f(\xi)h,
\eqan
where $\xi \in I_j$ and the last equality follows from the mean value theorem for continuous bounded functions. Intuitively, we can approximate the probability of $X$ falling into the interval $I_j$ by the proportion of observations in $I_j$, i.e., 
\bqan \label{eq.f_approx}
P(X \in I_j) \approx \frac{\#\{x_i \in I_j\}}{n}.
\eqan
\indent Using the approximation in (\ref{eq.f_approx}) and the equation in  (\ref{eq.f}), the density function $f(x)$ can be estimated by
\bqan\label{eq.f_histogram}
\hat{f}_h(x) = \frac{\#\{x_i \in I_j\}}{n h} = \frac{1}{n h}\sum_{i=1}^n \mathbbm{1}(x_i \in I_j) \mbox{    for  } x \in I_j,
\eqan
where 
\bqa
 \mathbbm{1}(x_i \in I_j)  = \left\{ \begin{array}{ll}
         1 &  \mbox{if }  x \in I_j\\
        0 & \mbox{otherwise.}\end{array} \right.
\eqa
\indent The smoothness of the histogram estimate is controlled by the smoothing parameter $h$, a characteristic shared by all nonparametric curve estimators. Choosing a small bandwidth leads to a jagged estimate, while larger bandwidths tend to produce over smoothed histogram estimates (see \citeasnoun{Hardle1990}). Figure \ref{Fig.hist} shows an example of two histograms of the same randomly generated data: the histogram on the left hand side was estimated with a small bandwidth and consequently has many bins, while the histogram on the right hand side was computed with a large bandwidth, producing a smaller number of bins. The choice of the bandwidth is discussed in more detail in Section \ref{sec.bandwidth}. Note that in practice, the choice of $k$ will determine $h$ or vice versa (a rule of thumb for the choice of $k$ is the Sturges'  rule: $k = 1 + \log_2 n$). 



\indent When building a histogram, not only the bandwidth needs to be chosen, but also the starting point of each bin edge. These choices can produce different impressions of the shape, and hence different estimates. The bin edge problem is a disadvantage of the histogram not shared by other estimators, such as the kernel density estimator. Another disadvantage is that the histogram estimators are usually not smooth, displaying bumps that may have been observed only due to noise. 

\section{Kernel Density Estimation} \label{sec.kernel}

\indent \indent In econometrics, kernel density estimation is also known as the Parzen-Rosenblatt window method. It is an approach that is rooted in the histogram methodology. The basic idea is to estimate the density function at a point $x$ using neighboring observations. However, instead of building up the estimate according to bin edges, the naive kernel method (adaptively) uses each point of estimation $x$ as the center of the bin of width $2h$. To express it more transparently, consider the weight function
\bqan \label{kernel.weight}
K(x)  = \left\{ \begin{array}{ll}
         \frac{1}{2} &  \mbox{if }  |x| < 1\\
        0 & \mbox{otherwise,}\end{array} \right.
\eqan
called the kernel weight. Then, the kernel estimate (\citeasnoun{Rosenblatt1956}) of $f(x)$ is defined as
\bqan\label{kernel.estimator}
\hat{f}(x) = \frac{1}{n h}\sum_{i=1}^nK\left(\frac{x - X_i}{h}\right).
\eqan
This kernel density estimator is specifically called naive because the kernel weight used is simply a bin of width $2h$ centered at $x$. See \citeasnoun{Silverman1986} for a deeper discussion about this kind of estimator. 

\indent Note that the estimator in (\ref{kernel.estimator}) is an additive function of the kernel weight, inheriting properties such as continuity and differentiability. Hence, it is not continuous and has zero derivatives everywhere except on the jump points $X_i \pm h$. Moreover, even with a good choice of $h$, estimators that use weights as in (\ref{kernel.weight}) most often do not produce reasonable estimates of smooth densities. This is because the discontinuity of the kernel weight gives the estimate function a ragged form, creating sometimes misleading impressions due to several bumps and constant estimates where few data points are observed. 
As an illustration, we consider the CEO compensation data in 2012, containing the 200 highest paid chief executives in the U.S. This data set can be obtained from the Forbes website http://www.forbes.com/lists/2012/12/ceo-compensation-12\_rank.html. For a better visualization of the plot, we excluded the number 1 in the ranking, with an income of US\$131.19 mil, as it was an outlier. 

%
\vspace{-.6cm}
\begin{center}
\begin{figure}[!htbp] 
	\centering
	\includegraphics[width=1.1\textwidth, angle=0]{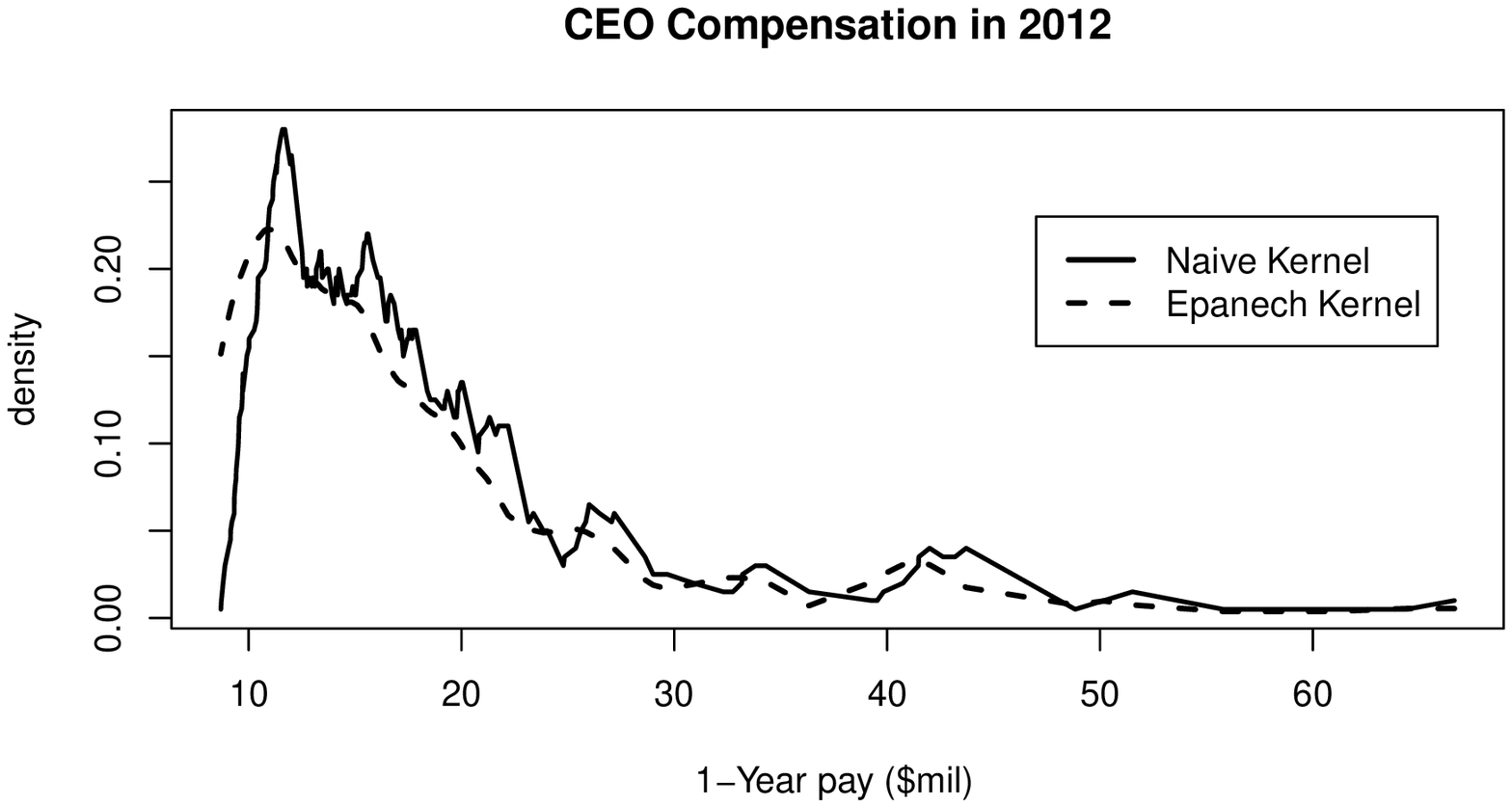}
	\caption{Estimated density of CEO compensation using the naive(solid line) and the Epanechnikov(dashed line) kernels. } \label{Fig.ragged}
\end{figure}
\end{center}
\vspace{-1.7cm}
Figure \ref{Fig.ragged} shows two density estimators: the solid line represents the naive estimator, while the dashed line represents a more adequate kernel type, called Epanechnikov, which will be described later. The density estimated by the naive kernel appears to have several small bumps, which are probably due to noise, not a characteristic of the true underlying density. On the other hand, the Epanechnikov kernel is smooth, avoiding this issue.

\indent A usual choice for the kernel weight $K$ is a function that satisfies
$
\int_{-\infty}^{\infty}K(x)dx = 1.
$
If moreover, it is assumed that $K$ is a unimodal probability density function that is symmetric about $0$, then the estimated density $\hat{f}(x)$ is guaranteed to be a density.
Note that the weight in (\ref{kernel.weight}) is an example of such choice. Suitable weight functions help overcome problems with bumps and discontinuity of the estimated density. For example, if $K$ is a gaussian distribution, the estimated density function $\hat{f}$ will be smooth and have derivatives of all orders. Table \ref{table.kernel_types} presents some of the most used kernel functions and Figure \ref{fig.kernel_types}  displays the format of the Epanechnikov, Uniform, Gaussian and Triweight kernels.



{\centering
\begin{minipage}{0.45\textwidth}
  \centering
\vspace{.6cm}
\begin{tabular}{l c} 
\hline\hline  
Kernel weight  & K(x)   \\ 
\hline
Uniform     & $\frac{1}{2}\mathbbm{1}(|x| < 1)$ \\
Gaussian  & $\frac{1}{\sqrt{2\pi}}e^{-\frac{1}{2}x^2}$ \\
Epanechnikov     & $\frac{3}{4}(1-x^2)\mathbbm{1}(|x| \leq 1)$ \\
Biweight     & $\frac{15}{16}(1-x^2)^2\mathbbm{1}(|x| \leq 1)$ \\
Triweight     & $\frac{35}{32}(1-x^2)^3\mathbbm{1}(|x| \leq 1)$ \\
\hline   
\end{tabular}
\vspace{.5cm}
\captionof{table}{Most common Kernel weights}\label{table.kernel_types}
\end{minipage}
\begin{minipage}{0.553\textwidth}
  \centering
  \includegraphics[width=\textwidth]{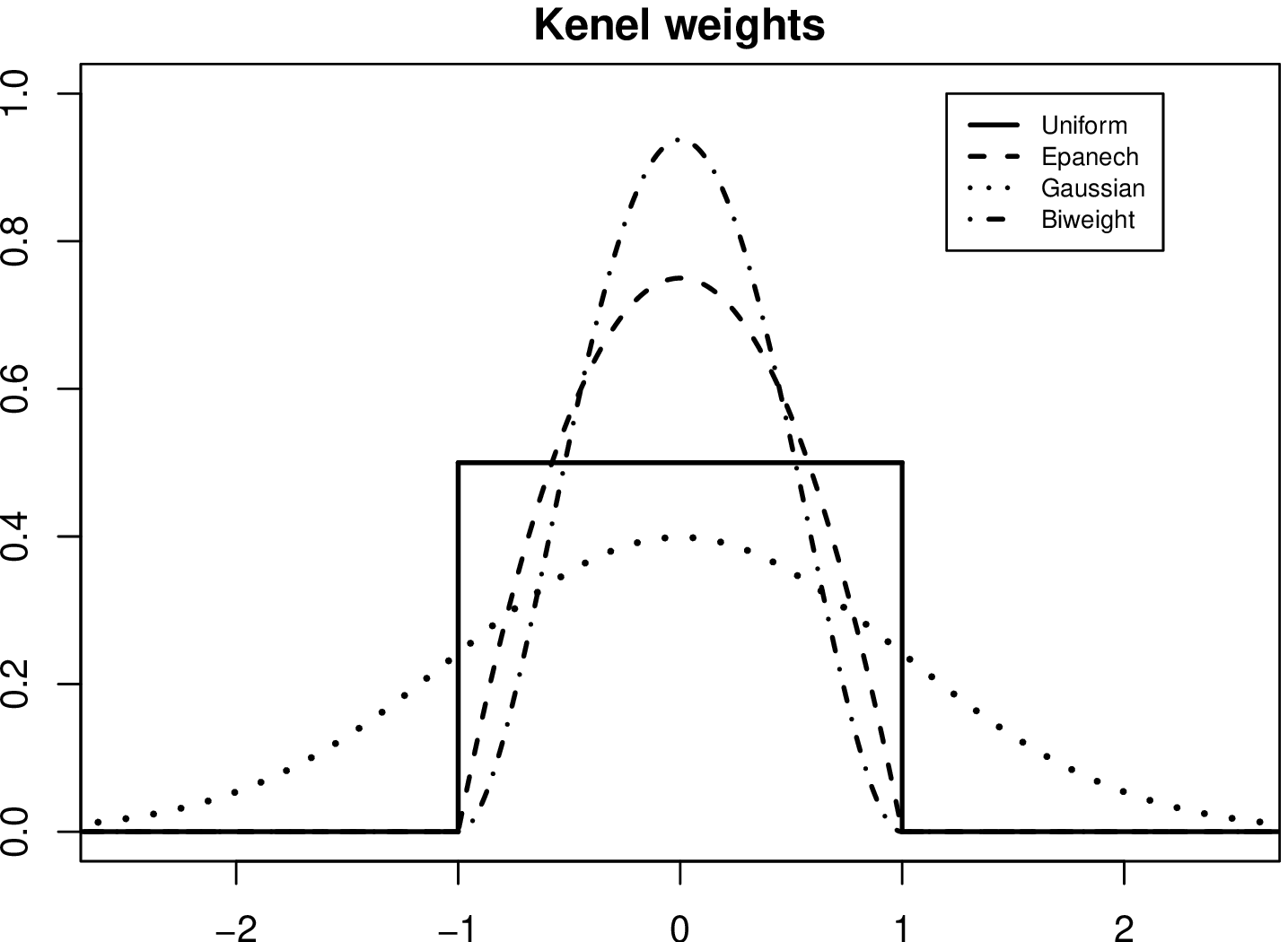}\label{fig.kernel_types}
  \captionof{figure}{Kernel weight functions}
\end{minipage}
}




One of the drawbacks of the kernel density estimation is that it is always biased, particularly near the boundaries (when the data is bounded). However, the main drawback of this approach happens when the underlying density has long tails. In this case, if the bandwidth is small, spurious noise appears in the tail of the estimates, or if the bandwidth is large enough to deal with the tails, important features of the main part in the distribution may be lost due to the over-smoothing. To avoid this problem, adaptive bandwidth methods have been proposed, where the size of the bandwidth depends on the location of the estimation. See Section \ref{sec.bandwidth} for more details on bandwidth selection.

\subsection{Properties of Kernel density estimators} \label{Sec.properties}

\indent \indent In this section, some of the theoretical properties of the kernel density estimator are derived, yielding reliable practical use. Assume we have $X_1, \ldots, X_n$ i.i.d. random variables from a density $f$ and let $K()$ be a Kernel weight function such that the following conditions hold
\bqa
\int K(u)du = 1, \mbox{                }
\int u K(u)du = 0,\mbox{       and       }
\int u^2 K(u)du = \mu_2(K)> 0.
\eqa

Then, for a non-random $h$, the expected value of $\hat{f}(x)$ is
\bqan\label{eq.expected}
E(\hat{f}(x)) &=& \frac{1}{n h} \sum_{i=1}^nE\left(K\left(\frac{x - X_i}{h}\right)\right) = \frac{1}{h}E\left(K\left(\frac{x - X_i}{h}\right)\right)\\
 &=& \frac{1}{h}\int K\left(\frac{x - u}{h}\right)f(u)du = \int K(y)f(x - yh)dy
\eqan
It is easy to see that $\hat{f}$ is an asymptotic unbiased estimator of the density, since $E(\hat{f}(x)) \rightarrow f(x)\int K(y)dy = f(x)$ when $h \rightarrow 0$. It is important to note that the bandwidth strongly depends on the sample size, so that when the sample size grows, the bandwidth tends to shrink. 

\indent Now, assume also that the second derivative $f''$ of the underlying density $f$ is absolutely continuous and square integrable. Then, expanding $f(x + yh)$ in a Taylor series about $x$ we have 
\bqa
f(x - yh) = f(x) - hyf'(x) + \frac{1}{2}h^2y^2f''(x) + o(h^2)
\eqa
Then, using the conditions imposed on the Kernel, the bias of the density estimator is
\bqan\label{eq.bias}
Bias(\hat{f}(x)) = \frac{h^2}{2}f''(x)\mu_2(K) + o(h^2)
\eqan

The variance of the estimated function can be calculated using steps similar to those in (\ref{eq.expected}):
\bqa
Var(\hat{f}(x)) &=& \frac{1}{nh}\int K^2(y)f(x - hy)dy - \frac{1}{n}\left(E(\hat{f}(x))\right)^2\\
&=& \frac{1}{nh}\int K^2(y)\{f(x) + o(1)\}dy = \frac{1}{n}\{f(x) + o(1)\}^2\\
&=& \frac{1}{nh}\int K^2(y)dy f(x) + o\left(\frac{1}{nh}\right)\\
&=& \frac{1}{nh}R(K) f(x) + o\left(\frac{1}{nh}\right),
\eqa
where $R(g) = \int g^2(y)dy$ for any square integrable function $g$. From the definition of Mean Square Error (MSE), we have 
\bqa
MSE(\hat{f}(x)) &=& \int(\hat{f}(x) - f(x))^2dx = Var(\hat{f}(x)) + Bias^2(\hat{f}(x)) \\
&=&  \frac{1}{nh}R(K) f(x) + \frac{h^4}{4}f''^2(x)\mu_2^2(K)  +  o\left(\frac{1}{nh}\right) + o(h^4)
\eqa 

\indent It is straightforward to see that, in order for the kernel density estimation to be consistent for the underlying density, two conditions on the bandwidth are needed as $n \rightarrow \infty$: $h \rightarrow 0 \mbox{  and  } nh \rightarrow \infty$.
When these two conditions hold, $MSE(\hat{f}(x)) \rightarrow 0$, and we have consistency. Moreover, the trade-off between bias and variance is controlled by the MSE, where decreasing bias leads to a very noise (large variance) estimate and decreasing variance yields over-smoothed estimates (large bias). As has already been pointed out, the smoothness of the estimate depends on the smoothing parameter $h$, which is chosen as a function of $n$. For the optimal asymptotic choice of $h$, a closed form expression can be obtained from minimizing the Mean Integrated Square Error (MISE). Integrating the MSE over the entire line, we find (\citeasnoun{Parzen1962})
\bqan\label{MISE}
MISE(\hat{f}) = E\int(\hat{f}(x) - f(x))^2dx = \frac{R(K)}{nh} + \frac{h^4 \mu_2^2(K)R(f'')}{4},
\eqan
and the bandwidth $h$ that minimizes MISE is then
\bqan\label{eq.band_mise}
h_{\mbox{MISE}} = \left(\frac{R(K)}{\mu_2^2(K)R(f'')}\right)^{1/5}n^{-1/5}.
\eqan
Using this optimal bandwidth, we have
\bqan\label{eq.min_mise}
\inf_{h>0}MISE(\hat{f}) \approx \frac{5}{4}\left[\mu_2^2(K)R^4(K)R(f'')\right]^{1/5}n^{-4/5}.
\eqan
A natural question is how to choose the kernel function $K$ to minimize (\ref{eq.min_mise}). Interestingly, if we restrict the choice to a proper density function, the minimizer is the Epanechnikov kernel, where  $\mu_2^2(K)R^4(K) = 3^4/5^6$.

\indent The problem with using the optimal bandwidth is that it depends on the unknown quantity $f''$, which measures the speed of fluctuations in the density $f$, i.e., the roughness of $f$. Many methods have been proposed to select a bandwidth that leads to good performance in the estimation, some of these are discussed in Section \ref{sec.bandwidth}.

\indent The asymptotic convergence of the kernel density estimator has been widely explored. \citeasnoun{BickelRosenblatt1973} showed that for sufficiently smooth $f$ and $K$, $\sup_\ell |\hat{f}(x) - f(x)|/\sqrt{f(x)}$, when normalized properly, has an extreme value limit distribution. 
The strong uniform convergence of $\hat{f}$
\bqan\label{eq.unif.consist}
\lim_{n\rightarrow \infty} \sup_{x}|\hat{f}(x) - f(x)| = 0 \mbox{    } a.e.
\eqan
has been studied extensively when the observations are independent or weakly dependent. \citeasnoun{Nadaraya1965} showed that if $K$ is of bounded variation and if $f$ is uniformly continuous, then (\ref{eq.unif.consist}) holds as long as $\sum_{m\geq1}e^{\gamma m h_n^2} < \infty$ for each $\gamma > 0$. Moreover, 
\citeasnoun{Stute1982} derives a law of the logarithm for the maximal deviation between a kernel density estimator and the true underlying density function,  \citeasnoun{GineGuillou2002} find rates for the strong uniform consistency of kernel density estimators and \citeasnoun{EinmahlMason2005} introduce a general method to prove uniform in bandwidth
consistency of kernel-type function estimators.
Other results on strong uniform convergence with different conditions can be found in several other papers, such as
\citeasnoun{Parzen1962}, \citeasnoun{Bhattacharya1967}, \citeasnoun{VanRyzin1969}, \citeasnoun{MooreYackel1977},\citeasnoun{Silverman1978} and \citeasnoun{DevroyeWagner1980} .

\section{The choice of the smoothing parameter $h$} \label{sec.bandwidth}

\indent \indent Selecting an appropriate bandwidth for a kernel density estimator is of crucial importance, and the purpose of the estimation may be an influential factor in the selection method. In many situations, it is sufficient to subjectively choose the smoothing parameter by looking at the density estimates produced by a range of bandwidths. One can start with a large bandwidth, and decrease the amount of smoothing until reaching a "reasonable"  density estimate. However, there are situations where several estimations are needed, and such an approach is impractical. An automatic procedure is essential when a large number of estimations are required as part of a more global analysis.

\indent The problem of selecting the smoothing parameter for kernel estimation has been explored by many authors, and no procedure has yet been considered the best in every situation. Automatic bandwidth selection methods can basically be divided in two categories: classical and plug-in. Plug-in methods refer to those that find a pilot estimate of $f$, sometimes using a pilot estimate of $h$, and "plug it in"  the estimation of MISE, computing the optimal bandwidth as in (\ref{eq.band_mise}). Classical methods, such as cross-validation, Mallow's $C_p$, AIC, etc, are basically extensions of methods used in parametric modeling. \citeasnoun{Loader1999} discusses the advantages and disadvantages of the plug-in and classical methods in more detail. Besides these two approaches, it is possible to find an estimate of $h$ based on a reference density. Next, we present in more detail the reference method and the most used automatic bandwidth selection procedures.

\subsection{Reference to a Distribution}
\indent \indent A natural way to overcome the problem of not knowing $f''$ is to choose a reference density for $f$, compute $f''$ and substitute it in (\ref{eq.band_mise}). For example, assume that the reference density is Gaussian, and a Gaussian kernel is used, then
\bqa
h_{\mbox{MISE}} = \left(\frac{R(K)}{\mu_2^2(K)R(f'')}\right)^{1/5}n^{-1/5} = \left[\frac{(2\sqrt{\pi})^{-1}}{\frac{3}{8}\pi^{-1/2}\sigma^{-5}}\right]^{1/5} n^{-1/5} = 1.06\sigma n^{-1/5}.
\eqa
By using an estimate of $\sigma$, one has a data-based estimate of the optimal bandwidth. In order to have an estimator that is more robust against outliers, the interquartile range $R$ can be used as a measure of the spread. This modified version can be written as
\bqa
\hat{h}_{\mbox{robust}}= 1.06 \min\left(\hat{\sigma},\frac{\hat{R}}{1.34}\right)n^{-1/5}.
\eqa

%
%

\begin{center}
\begin{figure}[!htbp] 
	\centering
	\includegraphics[width=.92\textwidth, angle=0]{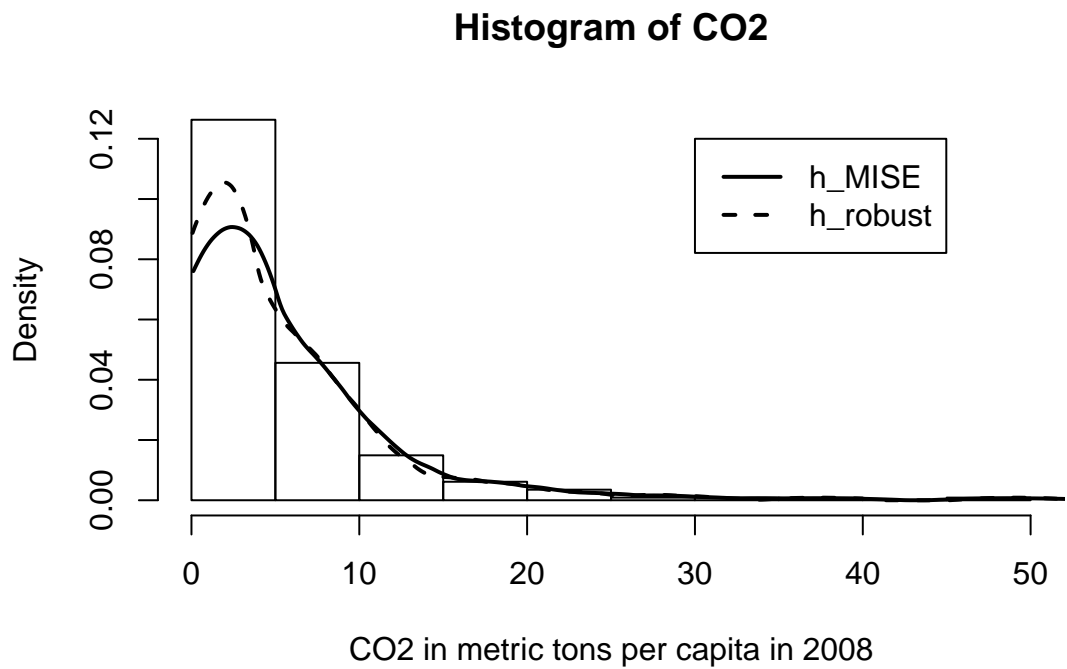}
	\caption{Estimated density of CO2 per capita in 2008 using the bandwidth that minimizes MISE(solid line) and the robust bandwidth(dashed line). } \label{Fig.ref_distr}
\end{figure}
\end{center}

Figure \ref{Fig.ref_distr} shows the estimated density of CO2 per capita in the year of 2008. The data set can be found at http://data.worldbank.org/indicator/EN.ATM.CO2E.PC/countries. Note that the estimated density that was computed with the robust bandwidth captures the peak that characterizes the mode, while the estimated density with the bandwidth that minimizes MISE smoothes out this peak. This happens because the outliers at the tail of the distribution contribute to $h_{MISE}$ be larger than the  robust bandwidth $h_{robust}$.
For more details on this estimator, see \citeasnoun{Silverman1986} or \citeasnoun{Hardle1990}.

\indent These methods are of limited practical use, since they are restricted to situations where a pre-specified family of densities is correctly selected. Plug-in and classical methods, described below, do not suffer from this limitation.

\subsection{Plug-in Methods}
\indent \indent There are several papers that address the plug-in approach for bandwidth selection. Some of them study different ways to estimate $R(f'')$, others explore ideas on how to select a pilot bandwidth to better estimate $R(f'')$. The idea is that the only unknown part of (\ref{eq.band_mise}) needs to be estimated, and hence the bandwidth estimator $h_{\mbox{MISE}}$ can be obtained. 

\indent \citeasnoun{ScottEtAl1977} proposed a sequential process: calculate $\hat{R}(f'') = R(\hat{f}_{h_2}(x))$, plug $\hat{R}(f'')$ into (\ref{eq.band_mise}) to obtain $h_3$, and iterate until convergence of the bandwidth. \citeasnoun{HallMarron1987} proposed estimating $\hat{R}(f^{(p)})$ by $\hat{R}(f_h^{(p)}) = R(\hat{f}_h^{(p)}) - \frac{R(K^{(p)})}{nh^{2p+1}}$. \citeasnoun{ParkMarron1990} modified this idea, estimating $\hat{R}(f^{(p)}) = R(\hat{f}_g^{(p)}) - \frac{R(K^{(p)})}{ng^{2p+1}}$ with $g$ having the optimal rate given in \citeasnoun{HallMarron1987}. An improvement of \citeasnoun{ParkMarron1990} method can be found in \citeasnoun{SeatherJones1991}. \citeasnoun{HallEtAl1991} proposed to use a kernel of order 2 and to take one extra term in the Taylor expansion of the integrated square bias, leading to
\bqan\label{eq.hall}
\mbox{MISE}_2(h) = \frac{R(K)}{nh} + \frac{h^4}{4}\mu_2^2(K)R(f'') - \frac{h^6}{24}\mu_2(K)\mu_4(K)R(f''').
\eqan
Since the minimizer of (\ref{eq.hall}) is not analytically feasible, they proposed to estimate the bandwidth by
\bqa
h_{\mbox{HSJM}} =   (\hat{J}_1/n)^{1/5} + \hat{J}_2(\hat{J}_1/n)^{3/5}
\eqa
where $\hat{J}_1 = \frac{R(K)}{\mu_2^2(K)\hat{R}(f'')}$ and $\hat{J}_2 = \frac{\mu_4(K)\hat{R}(f''')}{20\mu_2(K)\hat{R}(f'')}$.

\indent Several other plug-in methods have been proposed, and a review of the first procedures that address this type of methodology can be found in \citeasnoun{Turlach1993}. 
 Modern research on plug-in methods have actually become somewhat hybrid, combining ideas of plug-in and classical approaches such as cross validation, see Biased Cross-Validation described below for example.
 More recently, inspired by developments in threshold selection, \citeasnoun{ChanEtAl2010} propose to choose $h = O(n^{-1/5})$ as large as possible, so that the density estimator has a larger bias, but smaller variance than $\hat{f}_{h_{AMSE}(x)}$. The idea is to consider an alternative kernel density estimator $\bar{f}= \frac{1}{nh}\sum_{i=1}^n\bar{K}\left(\frac{x-X_i}{h}\right)$ and define $\Delta_n(x;h) = \frac{\sqrt{nh}\{\hat{f}(x;h) - \bar{f}(x;h)\}}{\hat{f}^{1/2}(x;h)\{\int (K(s) - \bar{K}(s))^2ds\}^{1/2}}$. Then, the choice for the smoothing parameter is 
\bqa
\hat{h}_l = arg\min\{h:|\Delta_n(x;r)| > z_\alpha \mbox{ for all } r > h, r\in[cn^{-1/5,n^{-\epsilon}}]\},
\eqa
where $z_\alpha$ denotes a critical point in $N(0,1)$, $c > 0$ and $0 < \epsilon < 1/5$. The intuition is that, when $h$ is large $\Delta_n(x;h) > z_\alpha$, since $\Delta_n(x;r) \dkonv N(0,1)$.

\subsection{Classical Methods}

\subsubsection{Least Squares Cross-Validation}

\indent \indent Cross-validation is a popular and readily implemented heuristic for selecting the smoothing parameter in kernel estimation. Introduced by \citeasnoun{Rudemo1982} and \citeasnoun{Bowman1984}, least squares cross-validation is very intuitive and has been a fundamental device in recent research. The idea is to consider the expansion of the Integrated Square Error (ISE) in the following way
\bqa
ISE(h) = \int \hat{f}_h^2(x)dx - \int \hat{f}_h(x)f(x)dx + \int f^2(x)dx.
\eqa
Note that the last term does not depend on $\hat{f}_h$, hence on $h$, so that we only need to consider the first two terms. The ideal choice of bandwidth is the one which minimizes
\bqa
L(h) = ISE(h) - \int f^2(x)dx = \int \hat{f}_h^2(x)dx - \int \hat{f}(x)f(x)dx.
\eqa
The principle of the least squares cross-validation method is to find an estimate of $L(h)$ from the data and minimize it over $h$. Consider the estimator
\bqan \label{eq.CV}
CV_{LS}(h) = \int \hat{f}_h^2(x)dx - 2\frac{1}{n}\sum_i \hat{f}_{h,-i}(X_i),
\eqan
where 
\bqan\label{eq.f_i}
\hat{f}_{h,-i}(X_i) = \frac{1}{(n-1)h}\sum_{j\neq i}K\left(\frac{x-X_j}{h}\right).
\eqan
The summation in (\ref{eq.CV}) has expectation
\bqa
E\frac{1}{n}\sum_i \hat{f}_{h,-i}(X_i) = E\hat{f}_{h,-n}(X_n) = E\int \hat{f}_{h,-n}(x)f(x)dx = E\int\hat{f}_h(x)f(x)dx,
\eqa
because $E(\hat{f}_h)$ depends only on the kernel and bandwidth, not on the sample size. It follows that $E(CV_{LS}(h)) = E(L(h))$, and hence $CV_{LS}(h) + \int f^2(x)dx$ is an unbiased estimator of MISE (reason why this method is also called {\it unbiased cross-validation}). Assuming that the minimizer of $CV_{LS}(h)$ is close to the minimizer of $E(CV_{LS}(h))$, the bandwidth
\bqa
h_{LSCV} = arg \min_h CV_{LS}(h)
\eqa
is the natural choice.  This method suffers from sample variation, that is, using different samples from the same distribution, the estimated bandwidths may have large variance. Further discussion on this method can be found in \citeasnoun{BowmanEtAl1984}, \citeasnoun{Hall1983} and \citeasnoun{Stone1984}.

\subsubsection{Biased Cross-Validation}
\indent \indent Biased cross-validation considers the asymptotic MISE
\bqa
AMISE\{\hat{f}_h\} = \frac{R(K)}{nh} + \frac{h^4}{4}\mu_2^2(K)R(f'').
\eqa
This method was suggested by \citeasnoun{ScottTerrell1987}, and its main idea is to replace the unknown quantity $R(f'')$ by the estimator
\bqa
\widetilde{R(f'')} &=& R(\hat{f}_h'') - (nh^5)^{-1}R(K'')\\
&=& n^{-2}\sum\sum_{i\neq j}(K_h''*K_h'')(X_i - X_j),
\eqa
to give
\bqa
BCV(h) = (nh)^{-1}R(K) + \frac{h^4}{4}\mu_2^2(K)\widetilde{R(f'')}.
\eqa
Then, the bandwidth selected is $h_{BCV} = arg \min_h BCV(h)$. This selector is considered a hybrid of cross-validation and plug-in, since it replaces an unknown value in AMISE by a cross-validatoin kernel estimate $\widetilde{R(f'')}$.

\subsubsection{Likelihood Cross-Validation}
\indent \indent Suppose that in addition to the original data set $X_1, \ldots, X_n$, we have another independent observation $X^*$ from $f$. Thinking of $\hat{f}_h$ as a parametric family depending on $h$, but with fixed data $X_1, \ldots, X_n$, we can view $\log\hat{f}(X^*)$ as the likelihood of the bandwidth $h$. Because in reality no additional observation is available, we can omit a randomly selected observation from the original data, say $X_i$, and compute $\hat{f}_{h,-i}(X_i)$, as in (\ref{eq.f_i}). Note that there is no pattern when choosing the observation to be omitted, so that the score function can be taken as the log likelihood average
\bqa
CV(h) = n^{-1}\sum_{i=1}^n\log \hat{f}_{h,-i}(X_i).
\eqa
Naturally, we choose the bandwidth the minimizes $CV(h)$, which is known to minimize the Kullback-Leibler distance between $\hat{f}_h(x)$ and $f(x)$. This method was proposed by \citeasnoun{HabbemaEtAl1974} and \citeasnoun{Duin1976}, but other results can be found in \citeasnoun{Marron1987}, \citeasnoun{Marron1989} and \citeasnoun{CaoEtAl1994}.

In general, bandwidths chosen via cross validation methods in kernel density estimation are highly variable, and usually give undersmooth density estimates, causing undesired spurious bumpiness.

\subsubsection{Indirect Cross-validation}
\indent \indent The Indirect Cross-validation (ICV) method, proposed by \citeasnoun{SavchukEtAl2010}, slightly outperforms least squares
cross-validation in terms of mean integrated squared error. The method can be described as follows. First define the family of kernels $\mathbb{L} = \{L(.; \alpha,\sigma):\alpha\geq0, \sigma > 0\}$ where, for all $u$, $L(u; \alpha,\sigma) = (1+\alpha)\phi(u) - \frac{\alpha}{\sigma}\phi\left(\frac{u}{\sigma}\right)$. Note that this is a linear combination of two gaussian kernels. Then, select the bandwidth of an L-kernel estimator using least squares cross-validation, and call it $\hat{b}_{UCV}$. Under some regularity conditions on the underlying density $f$, $h_n$ and $b_n$ that asymptotically minimize the MISE of $\phi$ and L-kernel estimators, have the following relation 
\bqa
h_n = \left(\frac{R(\phi)\mu_2^2(L)}{R(L)\mu_2^2(\phi)}\right)^{1/5}b_n = Cb_n.
\eqa
The indirect cross-validation bandwidth is chosen to be $\hat{h}_{ICV} = C\hat{b}_{UCV}$.  \citeasnoun{SavchukEtAl2010} show that the relative error of ICV bandwidths can converge to 0 at a rate of $n^{1/4}$, much better than the $n^{1/10}$ rate of LSCV.

\subsection{Other Methods}

\subsubsection{Variable Bandwidth}
\indent \indent Rather than using a single smoothing parameter $h$, some authors have considered the possibility of using a bandwidth $h(x)$ that varies according to the point $x$ at which $f$ is estimated. This is often referred as the {\it balloon} estimator and has the form
\bqan\label{balloon}
\hat{f}(x) = \frac{1}{n h(x)}\sum_{i=1}^nK\left(\frac{x - X_i}{h(x)}\right).
\eqan
The {\it balloon} estimator was introduced by \citeasnoun{LoftsgaardenQuesenberry1965} in the form of the k{\it th} nearest neighbor estimator. In \citeasnoun{LoftsgaardenQuesenberry1965}, $h(x)$ was based on a suitable number $k$, so that it was a measure of the distance between $x$ and the {\it k}th data point nearest to $x$. The optimal bandwidth for this case can be shown to be (analogue of (\ref{eq.band_mise}) for asymptotic MSE)
\bqan\label{eq.band_mise_x}
h_{\mbox{AMSE}}(x) = \left(\frac{R(K)f(x)}{\mu_2^2(K)f''^2(x)}\right)^{1/5}n^{-1/5}.
\eqan

Another variable bandwidth method is to have the bandwidth vary not with the point of estimation, but with each observed data point. This type of estimator, known as sample-point or variable kernel density estimator, was introduced by \citeasnoun{BreimanEtAl1977} and has the form
\bqan\label{variable_kernel}
\hat{f}(x) = \frac{1}{n h(X_i)}\sum_{i=1}^nK\left(\frac{x - X_i}{h(X_i)}\right).
\eqan
This type of estimator has one advantage over the {\it balloon} estimator: it will always integrate to 1, assuring that it is a density. Note that $h(X_i)$ is a function of random variables, and thus it is also random.

\indent More results on the variable bandwidth approach can be found in \citeasnoun{Hall1992}, \citeasnoun{TaronEtAl2005}, \citeasnoun{WuEtAl2007} and  \citeasnoun{GineHailin2010}.

\subsubsection{Binning}
\indent \indent An adaptive type of procedure is the binned kernel density estimation, studied by a few authors such as \citeasnoun{Scott1981}, \citeasnoun{Silverman1982} and \citeasnoun{Jones1989}. The idea is to consider equally spaced bins $B_i$ with centers at $t_i$ and bin counts $n_i$, and define the estimator as
\bqan\label{binnned}
\hat{f}_{bin}(x) = \frac{1}{n}\sum_{i=-\infty}^\infty n_iK\left(\frac{x-t_i}{h}\right) = \frac{1}{n}\sum_{i=1}^m K\left(\frac{x-t_i}{h}\right),
\eqan
where the sum over $m$ means summing over the finite non-empty bins that exist in practice. It is also possible to use a variable bandwidth in (\ref{binnned}), yielding the estimator
\bqan\label{binnned_tilde}
\tilde{f}_{bin}(x) = \frac{1}{n}\sum_{i=1}^m K\left(\frac{x-t_i}{h(X_i)}\right),
\eqan

Examples of other approaches and discussion on this type of estimation can be found in \citeasnoun{HallWand1996}, \citeasnoun{Cheng1997}, \citeasnoun{Minnotte1998}, \citeasnoun{PawlakStadtmuller1999}, \citeasnoun{Holmstrom2000}.

\subsubsection{Bootstrap}
\indent \indent A methodology that has been recently explored is that of selecting the bandwidth using bootstrap. It focuses on replacing the MSE by $MSE^*$, a bootstrapped version of MSE, which can be minimized directly. Some authors resample from a subsample of the data $X_1, \ldots, X_n$ (see \citeasnoun{Hall1990}), others replace from a pilot density based on the data (see \citeasnoun{FarawayJhun1990}, \citeasnoun{Hazelton1996}, \citeasnoun{Hazelton1999}), more precisely, from $\tilde{f}^b_h(x) = \frac{1}{nb_n}\sum_{i=1}^nL\left(\frac{x-X_i}{b_n}\right)$, where $L$ is another kernel and $b_n$ is a pilot bandwidth. Since the bandwidth choice reduces to estimating $s$ in $h = n^{-1/5}s$, \citeasnoun{Ziegler2006} 
introduces $f^*_{n,s}(x)= \frac{1}{n^{4/5}s}\sum_{i=1}^nK\left(\frac{x-X_i^*}{n^{-1/5}s}\right)$, and obtain $MSE^*_{n,s}(x) = E^*((f^*_{n,s}(x) - \tilde{f}^b_h(x) )^2)$. The proposed bandwidth is 
\bqa
h_n = n^{-1/5} arg\min_s MSE^*_{n,s}.
\eqa
\indent Applications of the bootstrap idea can be found in many different areas of estimation, see \citeasnoun{DelaigleGijbels2004}, \citeasnoun{LohJang2010} for example.

\subsubsection{Estimating Densities on $\mathbb{R}_+$}

\indent \indent It is known that kernel density estimators have larger bias on the boundaries. Many methods have been proposed to alleviate such problem, such as the use of gamma kernels or inverse and reciprocal inverse gaussian kernels, also known as varying kernel approach. \citeasnoun{Chen2000} proposes to replace the symmetric kernel by a gamma kernel, which has flexible shapes and locations on $\mathbb{R}_+$. Their estimator can be described in the following way. Suppose the underlying density $f$ has support $[0,\infty)$ and consider the gamma kernel
\bqa
K_{x/b + 1, b}(t) = \frac{t^{x/b}e^{-t/b}}{b^{x/b+1}\Gamma(x/b+1)},
\eqa
where $b$ is a smoothing parameter such that $b\rightarrow 0$ and $nb \rightarrow \infty$. Then, the gamma kernel estimator is defined as 
\bqa
\hat{f}^G(x) = \frac{1}{n}\sum_{i=1}^nK_{x/b+1,b}(X_i).
\eqa
\indent The expected value of this estimator is
\bqa
E\hat{f}^G(x) = \int_{0}^{\infty}K_{x/b + 1, b}(y)f(y)dy = Ef(\xi_x),
\eqa
where $\xi_{x}$ is a Gamma(x/b+1,b) random variable. Using Taylor Expansion and the fact that $E(\xi_x) = x + b$ and $Var(\xi_x) = xb + b^2$ we have that
\bqa
Ef(\xi_x) &=& f(x + b) + \frac{1}{2}f''(x)Var(\xi_x) + o(b)\\
&=& f(x) + b\left[f'(x) + \frac{1}{2}xf''(x)\right] + o(b).
\eqa
It is clear then, that this estimator does not have bias problems on the boundaries, since the bias is $O(b)$ near the origin and in the interior. See \citeasnoun{Chen2000} for further details. Other approaches on estimating the density on $\mathbb{R}_+$ can be found in \citeasnoun{Scaillet2004}, \citeasnoun{MnatsakanovRuymgaart2012}, \citeasnoun{MnatsakanovSarkisian2012}, \citeasnoun{ComteCatalot2012} and references therein.

\indent Some interest on density estimation research is on bias reduction techniques, which can be found in \citeasnoun{JonesEtAl1995}, \citeasnoun{ChoiHall1999}, \citeasnoun{ChengEtAl2000}, \citeasnoun{ChoiEtAl2000} and \citeasnoun{HallMinnotte2002}.
Other recent improvements and interesting applications of the kernel estimate can be found in \citeasnoun{Hirukawa2010},\citeasnoun{LiaoEtAl2010}, \citeasnoun{MatuszykEtAl2010}, \citeasnoun{MiaoEtAl2012}, \citeasnoun{ChuEtAl2012}, \citeasnoun{GolyandinaEtAl2012} and \citeasnoun{CaiEtAl2012} among many others.

\subsubsection{Estimating the distribution function $F(x)$}

\indent \indent It is not uncommon to find situations where it is desirable to estimate the distribution function $F(x)$ instead of the density function $f(x)$. A whole methodology known as kernel distribution function estimation (KDFE) has been explored since \citeasnoun{Nadaraya1964} introduced the estimator 
\bqa
\hat{F}_h(x) = \frac{1}{n}\sum_{i=1}^nK\left(\frac{x-X_i}{h}\right),
\eqa
where $K$ is the distribution function of a positive kernel $k$, i.e $K(x) = \int_{-\infty}^xk(t)dt$. Authors have considered many alternatives for this estimation, but the basic measures of quality or this type of estimator are
\bqa
ISE(h) &=& \int[\hat{F}_h(x) - F(x)]^2W(x)dF(x) \mbox{ and }\\
MISE(h) &=& E\int[\hat{F}_h(x) - F(x)]^2W(x)dF(x),
\eqa
where $W$ is a non-negative weight function.

\indent \citeasnoun{Sarda1993} considered a discrete approximation to MISE, the average squared error
\bqa
ASE(h) = \frac{1}{n}\sum_{i=1}^n[\hat{F}_h(X_i) - F(X_i)]^2W(X_i).
\eqa
He suggests replacing the unknown $F(X_i)$ by the empirical $F_n(X_i)$ and then selecting the bandwidth that minimizes the leave-one-out criterion
\bqa
CV(h) = \frac{1}{n}\sum_{i=1}^n[\hat{F}_{h,-i}(X_i) - F_n(X_i)]^2W(X_i).
\eqa

As an alternative to this cross-validation criterion, \citeasnoun{AltmanLeger1995} introduce a plug-in estimator of the asymptotically optimal bandwidth. There is a vast literature on estimating kernel distribution functions, for example \citeasnoun{BowmanEtAl1998}, \citeasnoun{Tenreiro2006}, \citeasnoun{AhmadAmezziane2007}, \citeasnoun{JanssenEtAl2007}, \citeasnoun{BergPolitis2009}, just to cite a few.

\subsection{Example of Bandwidth Selection Methods}
\indent \indent It is well known that plug-in bandwidth estimators tend to select larger bandwidths when compared to the classical estimators. They are usually tuned by arbitrary specification of pilot estimates and most often produce over smoothed results when the smoothing problem is difficult. On the other hand, smaller bandwidths tend to be selected by classical methods, producing under smoothed results. The goal of a selector of the smoothing parameter is to make that decision purely from the data, finding automatically which features are important and which should be smoothed away.

\begin{center}
\begin{figure}[!htbp] 
	\centering
	\includegraphics[width=\textwidth]{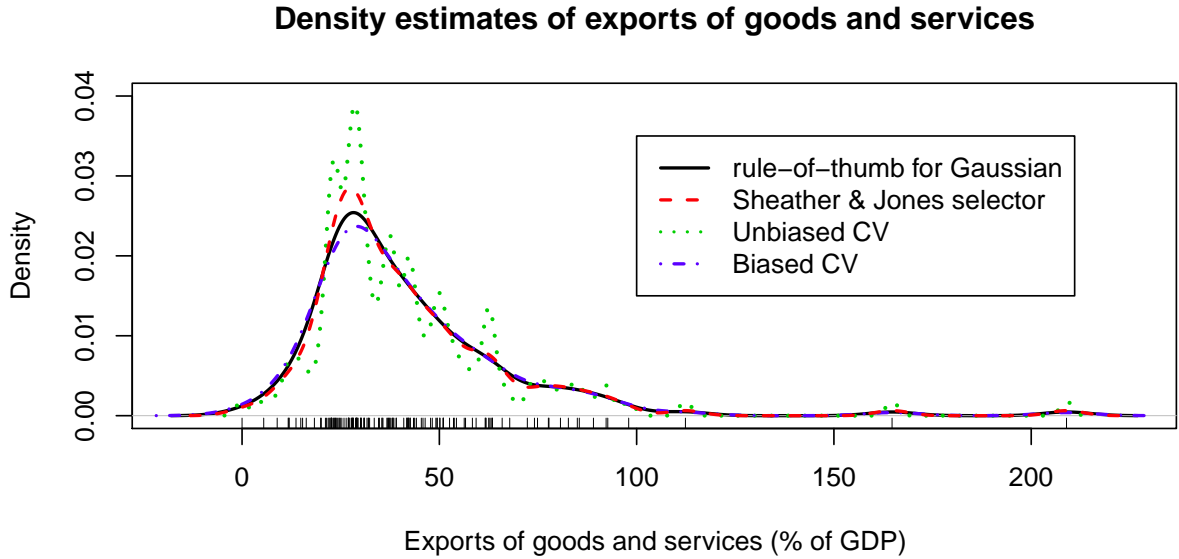}
	\caption{Estimated densities for bandwidths chosen using different methods.} \label{Fig.density_ex}
\end{figure}
\end{center}

\indent Figure \ref{Fig.density_ex} shows an example of classical and plug-in bandwidth selectors for a real data set. The data corresponds to the exports of goods and services of countries in 2011, representing the value of all goods and other market services provided to the rest of the world. The data set can be downloaded from the world bank website (http://data.worldbank.org). 



   
The plug-in estimators a) rule of thumb for Gaussian and b) Seather and Jones selector produced a very smooth fit, while unbiased cross-validation selects a small bandwidth, yielding a highly variable density estimate. The hybrid method biased cross-validation, is the one that selects the largest bandwidth, hence its corresponding density estimate is very smooth, smoothing away information of the peak (mode).
              
\section{SiZer} \label{sec.sizer}

\begin{center}
\begin{figure}[!htbp] 
	\centering
	\includegraphics[width=\textwidth]{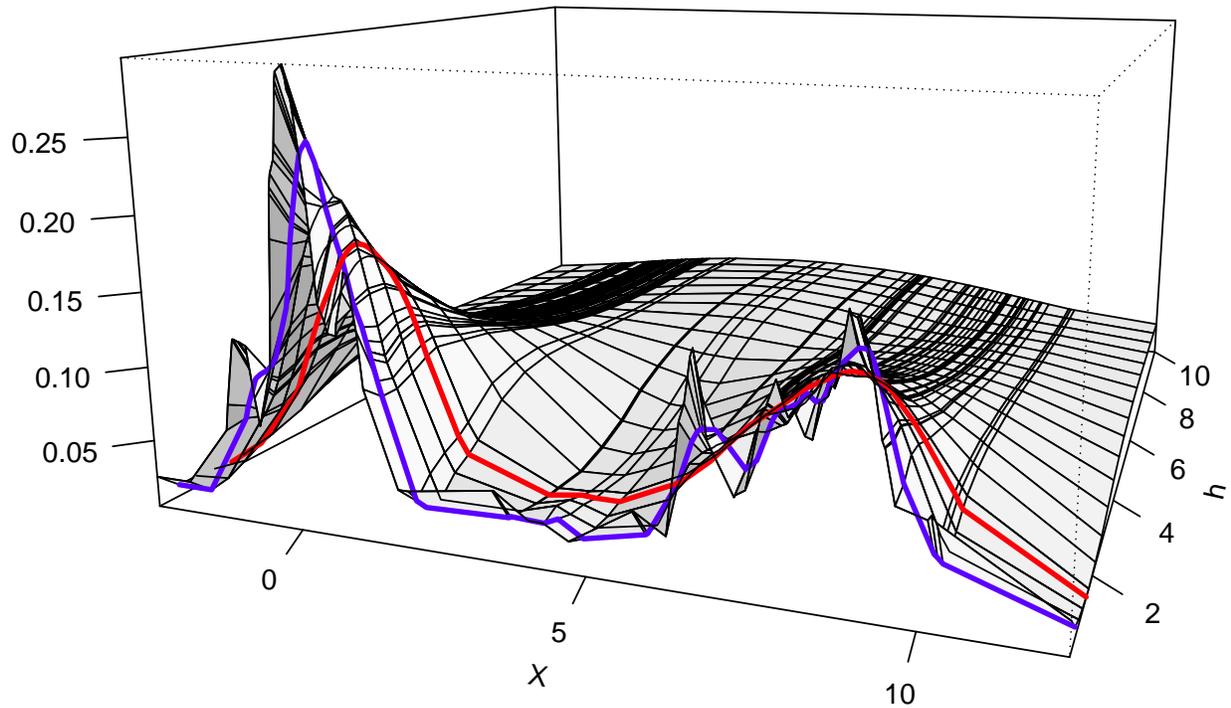}
	\caption{Estimated density with several bandwidths.} \label{Fig.mixture}
\end{figure}
\end{center}

\indent \indent In nonparametric estimation, the challenge of selecting the smoothing parameter that yields the best possible fit has been addressed through several methods, as described in previous sections. The challenge is to identify the features that are really there, but at the same time to avoid spurious noise.
 \citeasnoun{MarronChung1997} and other authors noted that it may be worth to consider a family of smooths with a broad range of bandwidths, instead of a single estimated function. Figure \ref{Fig.mixture} shows an example of a density generated from a mixture of a Gaussian variable with mean 0 and variance 1 and another Gaussian variable, with mean 8 and variance 2. The density was estimated with a Epanechnikov kernel using bandwidths that vary from 0.4 to 10. The wide range of smoothing considered, from a small bandwidth producing a wiggly estimate to a very large bandwidth yielding nearly the simple least squares fit, allows a contrast of estimated features at each level of smoothing. The two highlighted bandwidths are equal to 0.6209704 and 1.493644, corresponding to the choice of biased cross-validation (blue) and to Silverman's Ôrule of thumbÕ (red) (see Silverman, 1986) respectively .
%
%
%
%
%
%
%
%

\indent The idea of considering a family of smooths has its origins in scale space theory in computer science. A fundamental concept in such analysis is that it does not aim at estimating one true curve, but at recovering the significant aspects of the underlying function, since different levels of smoothing may reveal different intrinsic features. Exploring this concept in a statistical point of view, \citeasnoun{ChaudhuriMarron2000} introduced a procedure called {\bf SI}ignificance {\bf ZER}o crossings of smoothed estimates (SiZer), whose objective is to analyze the visible features representing important underlying structures for different bandwidths. Next, we briefly describe such method.

\indent Suppose that $h \in H$, where $H$ is a subinterval of $(0,\infty)$, and $x \in I$, where $I$ is a subinterval of $(-\infty,\infty)$. Then the family of smooth curves $\{\hat{f}_h(x)| h \in H, x \in I\}$ can be represented by a surface called {\it scale space surface}, which captures different structures of the curve under different levels of smoothing. Hence, the focus is really on $E(\hat{f}_h(x))$ as $h$ varies in $H$ and $x$ in $I$, which is called in \citeasnoun{ChaudhuriMarron2000} as "true curves viewed at different scales of resolution".

\indent A smooth curve $\hat{f}_h(x)$ has derivatives equal to 0 at points of minimum (valleys), maximum (peaks) and points of inflection. Note that, before a peak (or valley), the sign of the derivative $\partial \hat{f}_h(x)/\partial x$ is positive (or negative), and after it the derivative is negative (or positive). In other words, peaks and valleys are determined by zero crossings of the derivative. Actually, we can identify structures in a smooth curve by zero crossings of the m{\it th} order of the derivative. Using a Gaussian kernel $K(x) = (1/\sqrt{2\pi})exp(-x^2/2)$, \citeasnoun{Silverman1981} showed that the number of peaks in a kernel density estimate decreases monotonically with the increase of the bandwidth, and \citeasnoun{ChaudhuriMarron2000} extended this idea for the number of zero crossings of the {\it m}th order derivative $\partial^m \hat{f}_h(x)/\partial x^m$ in kernel regression.

\indent The asymptotic theory of the scale space surfaces and their derivatives studied by \citeasnoun{ChaudhuriMarron2000}, which hold even under bootstrapped or resampled distributions, provides tools for building bootstrap confidence intervals and tests of significance for their features (see \citeasnoun{ChaudhuriMarron1999}). SiZer basically considers the null hypothesis 
\bqa
H_0^{h,x}: \partial^m E(\hat{f}_h(x))/\partial x^m = 0,
\eqa
 for a fixed $x\in I$ and $h\in H$. If $H_0^{h,x}$ is rejected, there is evidence that $\partial^m E(\hat{f}_h(x))/\partial x^m$ is positive or negative, according to the sign of $\partial^m \hat{f}_h(x)/\partial x^m$.

\indent The information is displayed in a color map of scale space, where the pixels represent the location of x (horizontally) and h (vertically). The regions are shaded blue for significant increasing curve, red for significantly decreasing, purple for unable to distinguish and gray for insufficient data. Note that purple is displayed when the confidence interval for the derivative contains 0. There are a few options of software available, including java (http://www.wagner.com/SiZer/SiZerDownload.html), matlab (http://vm1.cas.unc.edu/stat-or/webspace/miscellaneous/marron/Matlab7Software/Smoothing/) and R (SiZer package).

\begin{center}
\begin{figure}[!htbp] 
	\centering
	\includegraphics[width=0.7\textwidth]{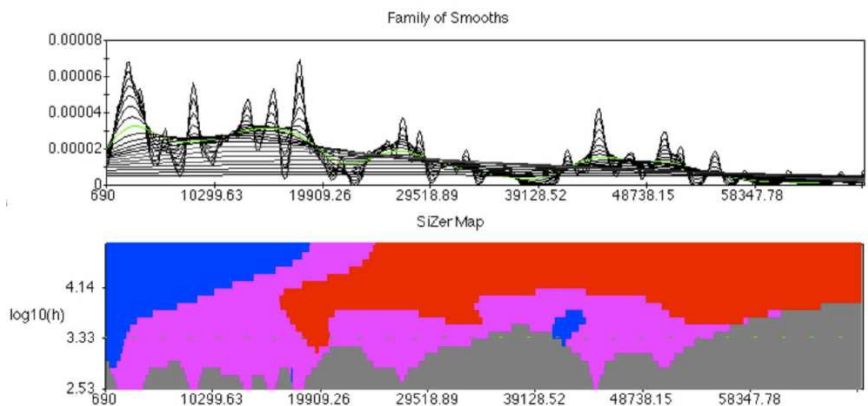}
	\caption{SiZer example} \label{Fig.employ2010}
\end{figure}
\end{center}

\indent Figure \ref{Fig.employ2010} shows an example of a color map obtained with SiZer. The data is the GDP per person employed in 2010 (downloadable at http://data.worldbank.org). It is easy to see that for large bandwidths, the density function significantly increases until about 16000, then after a small area that SiZer is unable to distinguish, it has a significant decrease, hence estimating a density with one mode at around 16000. Small bandwidths produce a map that is mostly gray, meaning that the wiggles in the estimate at that level of resolution
can not be separated from spurious sampling noise. An interesting blue area appears, with a mid-level resolution, near 43000, indicating a slightly significant increase. This comes after and before a purple area, which SiZer is unable to distinguish if it is increasing or decreasing. Thus, with a mid-level bandwidth, the estimated density would suggest 2 modes, one somewhere near 10000 and another near 43000.


\pagebreak 


{\bf Acknowledgments:} This paper was partially supported with grant 2012/10808-2 FAPESP (Funda\c{c}\~ao de Amparo \`a Pesquisa do Estado de S\~ao Paulo).

\bibliography{mybib}

\end{document}